\documentclass{article}
\usepackage{gdgspace,emlines2}
\advance\textheight by -\headsep\advance\textheight by -\headheight
\begin{document}
\def\cpw{{\it j}}
\def\aw{{\it i}}
\title{TSQT `Elements of Possibility'?}
\author{ R.E. Kastner\thanks{rkastner@wam.umd.edu}}
\date{December 10, 1998}
\maketitle
\begin{center}
{\small \em Department of Philosophy \\
University of Maryland \\
College Park, MD 20742 USA. \\}
\end{center}
\begin{abstract}
I defend my arguments in quant-ph/9806002, which
have recently been criticized by L. Vaidman (quant-ph/9811092).
I emphasize that the correct usage of the
ABL rule applies not to a genuine counterfactual statement
but rather to a conjunction of material conditionals.
I argue that the only kind of valid counterfactual
statement one can make using the ABL rule is
a ``might'' counterfactual, which is not adequate
for the attribution of `elements of reality' to a 
quantum system.
\end{abstract} 
\large
Lev Vaidman (1998) claims that I am ``distorting'' his assertions
(Kastner 1998). 
He suggests
that I am trying to ``find something beyond [these claims],''
i.e., about values of observables or hidden variables,
which he has not made. Yet the title of
a 1987 paper of which he is first author is:
``How to Ascertain the Values of $\sigma_x$, $\sigma_y$,
and $\sigma_z$ of a Spin-${1\over2}$ Particle'' 
(Vaidman, Aharonov, and Albert, 1987). Moreover, in
that article he refers to other ``curious
{\it properties}  of a quantum system within time intervals
between two experiments'' (my italics), as presented by Albert, Aharonov,
and D'Amato (1985). The latter paper,
as noted in my article, uses an explicitly {\it bona fide} 
(in my sense) counterfactual
reading of the ABL rule to support the claim that an
appropriately pre- and post-selected system could have
``definite, dispersion-free values'' 
of non-commuting sets of observables.
 These claims about dispersion-free values of observables
and curious properties of a system between measurements
are certainly ontological claims, and I believe I am
fully justified in treating them as such in my paper.
(However, it should be noted that none of
the arguments in my paper rely on a hidden variables 
interpretation of quantum mechanics, i.e., 
a modal interpretation such as Bohm's.) 

In any case, Vaidman's claim that what I identify
as the non-counterfactual reading of the ABL
rule---my ``first reading''---{\it is} in fact counterfactual, 
cannot be maintained. Here is the syntactical form
of that reading:\vskip .5cm
(If $P_1$ then $Q_1$) and (if $P_2$ then $Q_2$) and
 (if $P_3$ then $Q_3$) and (if....) and (if $P_N$ then $Q_N$),
\vskip .5cm
\noindent where each $P_i$ is ``A measurement of observable i
was performed'', and each $Q_i$ is ``the probability
of outcome $x_i$ was $p(x_i)$.''\vskip .5cm
This is clearly and unambiguously a conjunction
of material conditionals, not a counterfactual
statement. Applying this to a measurement context,
if only one of the N observables---say observable
$k$---is measured, then 
only one of the material conditionals has
a true antecedent; all the others have false antecedents
and are only vacuously true. As Mermin (1997) says in regard to a list
of possible outcomes obtained from the ABL rule 
for a time-symmetrized system, the 
outcomes on the list corresponding to conjuncts
with false antecedents are ``rubbish--they have nothing
to do with anything.''\footnote{Mermin (1997, p. 151).} 

Nevertheless, Vaidman wishes to call the other outcomes
``elements of reality,'' so he evidently does not
consider them rubbish. The argument apparently is that, by 
``fixing'' the system's two-state vector, the
other conjuncts can be considered applicable
to the same system. However, the only
way it can make any sense to ``hold
fixed'' the system's states at $t_1$ and $t_2$ 
is to talk about some {\it other} 
possible (i.e., non-actual) world
\cpw \ in which the counterfactual measurement
{\it is} performed and the system {\it happens} to
end up with the same two-state vector as in \aw.
But then all we are saying is that
there is {\it some} possible antecedent-world in which
the system has
the same two-state vector; we can make no
claim that the two-state vector is 
fixed
for {\it all} the closest possible antecedent-worlds,
as required . \footnote{That is, unless we arbitrarily designate
all worlds with the required two-state 
vector as consituting all 
the closest antecedent-worlds. (Vaidman has
indicated in a private correspondence that
this is in fact what he has in mind.) But, to borrow
a term from Bennett, this is 
certainly ``{\it ad hoc}
gerrymandering'' (1984, p. 62). Similarity of worlds
is properly based on comparative possibility,
not the accidental 
similarity of individual facts
such as having the same two-state vector;
this is not classical dogma but a very general 
and intuitive way to understand similarity of possible worlds.
(See Lewis 1973, e.g. p. 52).  Finkelstein (1998)
 also does not consider the antecedent-worlds 
in which the system has the same outcome at $t_2$
as in the
actual world to be closer to the actual world
than the antecedent-worlds in which the $t_2$ 
outcome differs. Moreover, Kvart's analysis 
of cotenability as requiring a lack of negative causal relevance
between the antecedent and background conditions
supports my assertion that ``fixing'' the two-state
vector in order to obtain a ``would''
counterfactual violates cotenability (see Kvart 1986, 
e.g. p.xii). }
 
In this case, the only kind of
counterfactual statement one can make 
is a ``might'' counterfactual rather than a ``would''
counterfactual. This is a statement like 
\vskip .5cm
If I were to strike this match, it might light.
\vskip .5cm
For the `might' counterfactual, it is not required that the
necessary background conditions hold throughout
the closest antecedent-permitting sphere, but merely at
{\it some} possible world in the closest 
antecedent-permitting sphere.
But this, of course, is a much weaker, essentially empty counterfactual.
Outcomes obtained from such a `might' counterfactual
would not merit the term ``elements of reality'';
the most one could say is that they are
 ``elements of possibility.''\footnote{This notion
of ``element of possibility'', however, is very weak.
For instance, it would appear to have
 no special significance as regards to
dispositional properties of a system. }
  
Finally, I would note that Vaidman's 
  ``...most convincing example that the term counterfactual
is appropriate'', referring to an example discussed in
his (1997), was also
discussed, but rather differently,
 in his (1996). In that paper, he notes (p.903)
that ``we cannot see this `reality' for one particle because the
uncertainty of 
the appropriate weak measurement has to be much larger than 
1...'',
but in his reply (1998) and in his (1997) he seems to suggest that the weak-measurement
 elements of
reality apply to a {\it single} particle. This point is
exactly what is under dispute.
\vskip .3cm
\noindent Acknowledgements\vskip .2cm
I would like to thank Jeffrey Bub, Jerry Finkelstein,
Charles Hagelgans, 
Jim Malley, and Lev Vaidman for useful discussions
and/or correspondence.
\vskip .3cm
References\vskip .2cm
\noindent Albert, D.Z., Aharonov, Y., and D'Amato, S. (1985), 
{\it Physical Review Letters 54}, 5.\newline
Bennett, J. (1984), `Counterfactuals and Temporal Direction,'
{\it The Philosophical Review XCIII}, p. 57.\newline
Finkelstein, J. (1998), `Space-Time Counterfactuals,'
Los Alamos preprint, quant-ph/9811057. \newline
Kastner, R.E. (1998), `Time-Symmetrized Quantum Theory,
Counterfactuals, and ``Advanced Action''', forthcoming
in {\it
Studies in History and Philosophy of Modern Physics},
Los Alamos preprint quant-ph/9806002.\newline
Kvart, I. (1986), {\it A Theory of Counterfactuals},
Indianapolis: Hackett Publishing Co.\newline
Lewis, D. (1973), {\it Counterfactuals}, Cambridge: 
Harvard University Press. \newline 
Mermin, N. D. (1997), `How to Ascertain the Values of 
Every Member of a Set of Observables That Cannot All 
Have Values,' in R. S. Cohen et al. (eds), 
{\it Potentiality, Entanglement and Passion-at-a-Distance}, 
149-157, Kluwer Academic Publishers.\newline
Vaidman, L. (1996),`Weak-Measurement Elements of Reality,'
{\it Foundations of Physics 26}, p.895.\newline
Vaidman, L. (1997), `Time Symmetrized Counterfactuals
in Quantum Theory,' Los Alamos preprint quant-ph/9807075.\newline
Vaidman, L. (1998), `Defending Time-Symmetrized Quantum
Counterfactuals,' Los Alamos preprint quant-ph/9811092.\newline
Vaidman, Aharonov, and Albert (1987), 
`How to Ascertain the Values of $\sigma_x$, $\sigma_y$, and
$\sigma_z$ pf a Spin-${1\over 2}$ Particle,'
{\it Phys Rev Lett 58}, 1385.  
\end{document}